\documentclass[fleqn,10pt]{wlscirep}
\usepackage[utf8]{inputenc}
\usepackage[T1]{fontenc}
\usepackage{graphicx} 
\usepackage{multirow}
\usepackage{array}
\title{Inductive transfer learning from regression to classification in ECG analysis}

\author[1,+,*]{Ridma Jayasundara}
\author[1,+]{Ishan Fernando}
\author[1,+]{Adeepa Fernando}
\author[1]{Roshan Ragel}
\author[2]{Vajira Thambawita}
\author[1,*]{Isuru Nawinne}

\affil[1]{Department of Computer Engineering, University of Peradeniya,
Peradeniya, Sri Lanka}
\affil[2]{SimulaMet, Oslo, Norway}

\affil[*]{corresponding authors : Ridma Jayasundara (ridmajayasundara@eng.pdn.ac.lk), Isuru Nawinne (isurunawinne@eng.pdn.ac.lk)}

\affil[+]{these authors contributed equally to this work}

\begin{abstract}
Cardiovascular diseases (CVDs) are the leading cause of mortality worldwide, accounting for over $30\%$ of global deaths according to the World Health Organization (WHO). Importantly, one-third of these deaths are preventable with timely and accurate diagnosis. The electrocardiogram (ECG), a non-invasive method for recording the electrical activity of the heart, is crucial for diagnosing CVDs. However, privacy concerns surrounding the use of patient ECG data in research have spurred interest in synthetic data, which preserves the statistical properties of real data without compromising patient confidentiality. This study explores the potential of synthetic ECG data for training deep learning models from regression to classification tasks and evaluates the feasibility of transfer learning to enhance classification performance on real ECG data. We experimented with popular deep learning models to predict four key cardiac parameters, namely, Heart Rate (HR), PR interval, QT interval, and QRS complex—using separate regression models. Subsequently, we leveraged these regression models for transfer learning to perform 5-class ECG signal classification. Our experiments systematically investigate whether transfer learning from regression to classification is viable, enabling better utilization of diverse open-access and synthetic ECG datasets. Our findings demonstrate that transfer learning from regression to classification improves classification performance, highlighting its potential to maximize the utility of available data and advance deep learning applications in this domain.
\end{abstract}
\begin{document}


\flushbottom
\maketitle

\thispagestyle{empty}

\section{Introduction}
Cardiovascular diseases (CVDs) refer to a group of conditions that affect the heart and blood vessels, impacting either the structure or function of the heart. The World Health Organization states that CVDs are the primary cause of death worldwide~\cite{rid1}. Early diagnosis of cardiac diseases is essential for improving treatment outcomes. Traditionally, the diagnosis of CVDs relies on the patient's medical history and clinical evaluations~\cite{rid2}.
Standard methods for examining hearts include echocardiography, cardiac magnetic resonance, and tubular angiography. However, these techniques are complex, time-consuming, and expensive. In contrast, an electrocardiogram (ECG) provides detailed cardiac information and is a simple, quick, and non-invasive method for detecting CVDs~\cite{rid5}. Frequently used by cardiologists, an ECG monitors heart activity, identifies irregular arrhythmic events~\cite{rid7}, assesses heart rate variability, forecasts myocardial infarction~\cite{rid9}, and screens for contractile dysfunction~\cite{rid10}.

In many instances, traditional diagnostic methods prove inefficient due to the need for complex manual analysis and highly trained medical experts to attain sufficient accuracy~\cite{rid6}. Additionally, these methods struggle to handle large amounts of diverse information efficiently.
This necessitates the development of dependable and straightforward methods for automated monitoring and diagnosis, leading to the creation of computer-aided diagnosis systems (CADS).

Deep learning (DL) algorithms, a branch of machine learning (ML), have emerged as powerful tools for developing robust models that can identify relationships within data and uncover hidden patterns in ECG readings. These models are particularly good at handling large datasets and typically improve as more data is incorporated, often surpassing traditional ML methods in performance~\cite{rid11, rid12}.

Convolutional neural networks (CNNs), initially designed for object recognition and image classification, are among the most frequently used DL algorithms~\cite{rid13}. While CNNs were first applied to 2 Dimensional (2D) data due to their ability to detect spatial hierarchies in images through convolutional layers, their powerful pattern recognition capabilities have since been adapted to one-dimensional (1D) data, such as time series. In this context, 1D CNNs process temporal sequences by applying convolutions along one dimension, enabling the detection of local patterns over time, which is particularly useful in ECG signal analysis.

A major challenge in developing accurate ECG classification models is the scarcity of available labeled data. While there are numerous ECG waveform datasets, many lack the classification labels necessary for training machine learning models. In addition, some datasets are limited to regression labels or contain unlabeled data altogether. This scarcity of labeled data poses a barrier to developing accurate and reliable ECG classification systems. For instance, a study on patient-specific ECG classification highlights this issue, noting that the limited availability of patient-specific data is a significant obstacle for training deep learning models effectively~\cite{rid21}. Another issue stems from privacy concerns, which restrict the free sharing of real ECG datasets~\cite{b1}. These concerns create additional barriers as researchers often struggle to obtain sufficient patient-specific data for training deep learning models. To address these challenges, we propose using synthetic ECG datasets, which contain regression labels which are extracted from ~\cite{b1}. This approach not only alleviates the privacy concerns associated with real data but also provides an indirect solution to the lack of manually labeled ECG data. 



\begin{figure}[h!]
    \centering
    \includegraphics[width=0.40\textwidth]{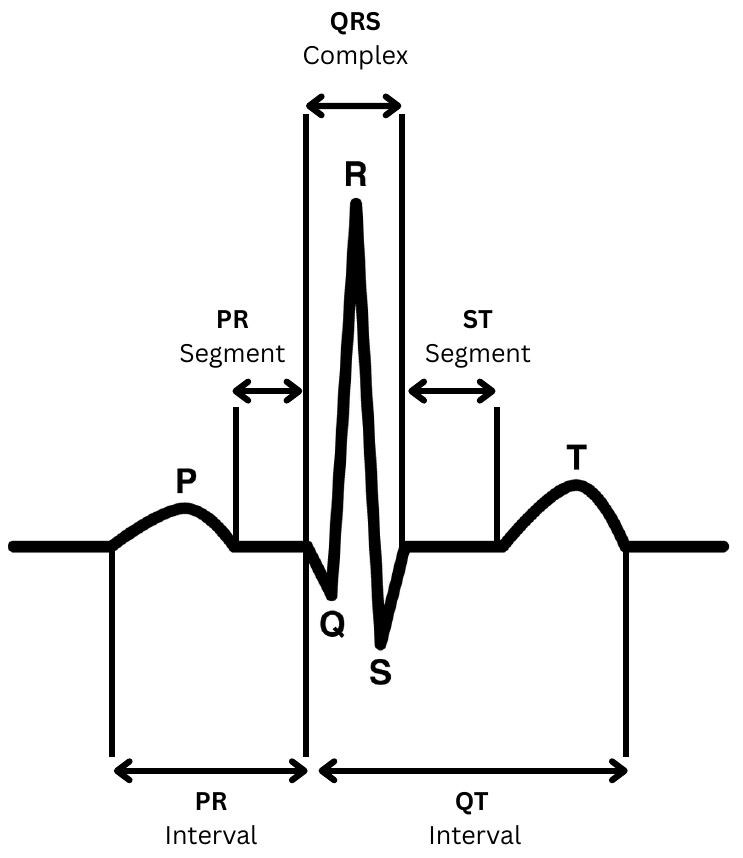} 
    \caption{Illustration of a typical ECG signal waveform highlighting key components of the cardiac cycle. The waveform includes the P wave (atrial depolarization), the QRS complex (ventricular depolarization), and the T wave (ventricular repolarization).\cite{b9}}
    \label{fig:ecg_signal_image}
\end{figure}

The electrocardiogram (ECG) shown in Figure~\ref{fig:ecg_signal_image} provides a visual representation of the heart’s electrical activity, which is crucial for diagnosing various cardiac conditions. It features distinct waves– P, Q, R, S, and T– each indicative of different phases of the cardiac cycle. The P wave signifies atrial depolarization, initiating atrial contraction, while the QRS complex represents ventricular depolarization, essential for ventricular contraction. The T wave, on the other hand, signifies ventricular repolarization, preparing the heart for the next cycle. Moreover, the ECG delineates intervals and segments: the PR interval measures atrioventricular conduction time, the QT interval monitors ventricular depolarization and repolarization, the PR segment reflects AV nodal conduction, and the ST segment indicates early ventricular repolarization. Changes in these intervals and segments can signify abnormalities like arrhythmias, ischemic heart disease, or conduction disorders. These ECG features can be extracted using the MUSE system automatically. 

In this study, we investigate the practical feasibility of applying inductive transfer learning—specifically, transferring knowledge from a regression task to a classification task—using both real and synthetic electrocardiogram (ECG) data. The main contributions of this study are:
\begin{itemize}
    \item Demonstration of the feasibility and efficacy of transferring knowledge from regression tasks to classification tasks in ECG analysis using deep learning models.
    \item Evaluation of the potential and effectiveness of synthetic data for transfer learning deep learning models, from regression to classification.
    \item  To the best of our knowledge, this is the first study to explore cross-domain transfer learning from synthetic to real ECG data, where the synthetic data is used for regression and the real data for classification.
    \item All the source code related to the experiments conducted in this study is available at our open-source GitHub to reproduce the results: \href{https://github.com/cepdnaclk/e18-4yp-GPU-Acceleration-for-Deep-Learning-based-Comprehensive-ECG-analysis}{repository}.
\end{itemize}


\section{Literature review}
One of the main goals of ECG analysis is to predict essential cardiac parameters such as heart rate, QT interval, QRS duration, and PR segment changes. These predictions are crucial for diagnosing cardiac conditions, assessing cardiovascular risk, and guiding treatment. Machine learning models, especially deep learning architectures, have demonstrated success in extracting complex patterns from ECG data to make accurate predictions~\cite{i1,i1_}.

Current applications of machine learning in ECG analysis include early detection of arrhythmias~\cite{r_early_ar} and personalized risk stratification~\cite{r_per_risk}, which can significantly enhance patient care and outcomes. The application of machine learning models in ECG analysis has seen significant advancements, with deep learning methods notably outperforming traditional signal processing techniques~\cite{r_tradi_sig}. Key studies in this domain highlight various models and their effectiveness. For instance, a Residual CNN applied to classify six types of abnormalities using the Telehealth Network of Minas Gerais, Brazil dataset~\cite{rid_tnmg} achieved a high specificity of $99\%$~\cite{b2}. Similarly, EfficientNet CNN~\cite{rid_effi} was utilized for diagnosing multilabel abnormalities, offering improved interpretability through Grad-CAM~\cite{rid_grad}, also leveraging the TNMG Brazil dataset. Echo State Networks (ESNs) demonstrated high accuracy and computational efficiency in heartbeat classification and arrhythmia detection through ensembles of ESNs~\cite{r_esn}.

Recurrent Neural Networks (RNNs)~\cite{ridRnn} have been comprehensively reviewed, with Long Short-Term Memory (LSTM) networks favored over other RNN architectures, achieving $88.1\%$ accuracy in binary classification with three hidden layers~\cite{b11}. A 1D CNN architecture featuring seven convolutional layers and three fully connected layers reached an accuracy of $86\%$ in ECG data classification. Deep Active Learning (DAL) with CNN produced high accuracy values, up to $99.02\%$, on the MIT-BIH dataset~\cite{b12}. An SVM (Support Vector Machines)~\cite{rid_svm} model also achieved an accuracy of $97.48\%$ for multiclass arrhythmia detection on the same dataset~\cite{b12}.  Attention-based models have been explored for ECG segmentation and classification, with a Conv-BiLSTM model leveraging self-attention to achieve high accuracy in ECG segmentation and fiducial point detection. Research into Transformer models~\cite{rid_attention}, including Wide and Deep Transformer Neural Networks~\cite{i1} and Heartbeat-aware Transformer models~\cite{r_h_aware}, showed strong performance in ECG classification, particularly for low data volumes and external validations. Vision Transformers~\cite{rid_vit}, such as ECG-Convolution-Vision Transformer Networks (ECVT-Net)~\cite{r_ecvt} and Vision Transformers with Deformable Attention, have outperformed traditional models like ResNet and LSTM, achieving high accuracy and F1 scores~\cite{b20}. These studies highlight the growing role of advanced machine learning techniques, particularly deep learning, in enhancing ECG analysis and classification.

Despite these advancements in model architectures, a significant challenge remains in the field of ECG analysis, which is the scarcity of available ECG data. This is mainly due to privacy concerns~\cite{b1, r_privacy} associated with sharing patient data. With the implementation of the General Data Protection Regulation (GDPR) in the European Union and  Personal Information Protection Law (PIPL) in China, the sharing of patient data has become increasingly restricted~\cite{b1}. Given these restrictions, generating synthetic medical data~\cite{r_generate} offers a promising solution. Properly generated synthetic data retains the statistical and clinical characteristics of real patient data without containing any personal information, making it impossible to trace back to an individual. This capability allows for the free exchange of data between research groups while maintaining compliance with stringent privacy regulations. Synthetic ECG data can thus provide a valuable resource for researchers, enabling them to develop and test algorithms without compromising patient confidentiality.

Synthetic data has been found to improve classification performance in a model designed to predict ventricular arrhythmias by enhancing the generalization of machine learning classifiers~\cite{rid_synthetic_2}. Among the various methods for generating synthetic data, generative adversarial networks (GANs)~\cite{r_gan} have emerged as a particularly promising approach. Studies have shown that augmenting datasets with synthetic ECGs improves performance in classification tasks~\cite{r_augment1,r_augment2}. For instance, a study utilizing BiLSTM-DC GANs demonstrated how augmenting real ECG data with synthetic signals enhances model accuracy for arrhythmia classification~\cite{rid_synthetic_1}. 

Further advancements in this field have been made using a modified WaveGAN~\cite{rid_waveGAN}, named WaveGAN*, to generate eight-channel ECG data~\cite{b1}. They also introduced a novel DeepFake ECG U-Net\cite{rid_u_net} model called Pulse2Pulse, specifically designed for generating 12-lead synthetic ECGs. This work provides a significant leap forward, demonstrating how synthetic ECGs can be used to preserve privacy while allowing open access to data for research~\cite{b1}.

While synthetic data generation provides a solution to data scarcity, the absence of classification labels remains a challenge. Transfer learning offers another promising approach to address this issue, allowing knowledge gained from one domain to be applied to a different but related domain. This approach is particularly valuable in scenarios with limited labeled data, which is common in medical applications such as ECG analysis. Transfer learning enables models to leverage patterns learned from data-rich source tasks to improve performance on target tasks with limited data availability, effectively addressing the challenges posed by data scarcity ~\cite{ishan_lit_review_1}.

In the context of deep learning, transfer learning typically involves pre-training a neural network on a large dataset (source domain) and then fine-tuning it on a smaller, specialized dataset (target domain). This methodology has shown remarkable success in various domains, including computer vision and natural language processing, before being applied to medical signal analysis~\cite{ishan_lit_review_2,rid_transfer}.

Inductive transfer learning~\cite{rid_inductive}, a specific type of transfer learning, assumes that while the source and target domains may have different data distributions, they share the same feature space. In this approach, labeled data is available for both domains, though typically in much larger quantities for the source domain. The model leverages the knowledge acquired from the source domain to improve performance in the target domain ~\cite{ishan_lit_review_3}.

Raghu et al. investigated the transferability of features in medical imaging and found that even simple transfer learning strategies can yield significant improvements over training from scratch~\cite{ishan_lit_review_4}. Similarly, Kachuee et al. applied transfer learning to ECG classification tasks and reported enhanced performance, particularly when the target dataset was limited in size~\cite{ishan_lit_review_5}.

In the specific context of ECG signal processing, Weimann and Conrad demonstrated how pre-training on large ECG datasets followed by fine-tuning on specific cardiac conditions could significantly improve diagnostic accuracy and reduce the need for extensive labeled data~\cite{ishan_lit_review_7}.

Research has also explored cross-domain transfer learning, where models trained on one type of physiological signal are adapted to analyze ECG data. Zhu et al. showed that features learned from EEG signals could be successfully transferred to ECG classification tasks, indicating the potential for knowledge sharing across different biomedical signal domains~\cite{ishan_lit_review_8}. Building on these insights, our research aims to explore a novel regression-to-classification transfer learning methodology for ECG analysis. 


\section{Methods}

\begin{figure}
    \centering
    \includegraphics[width=1\linewidth]{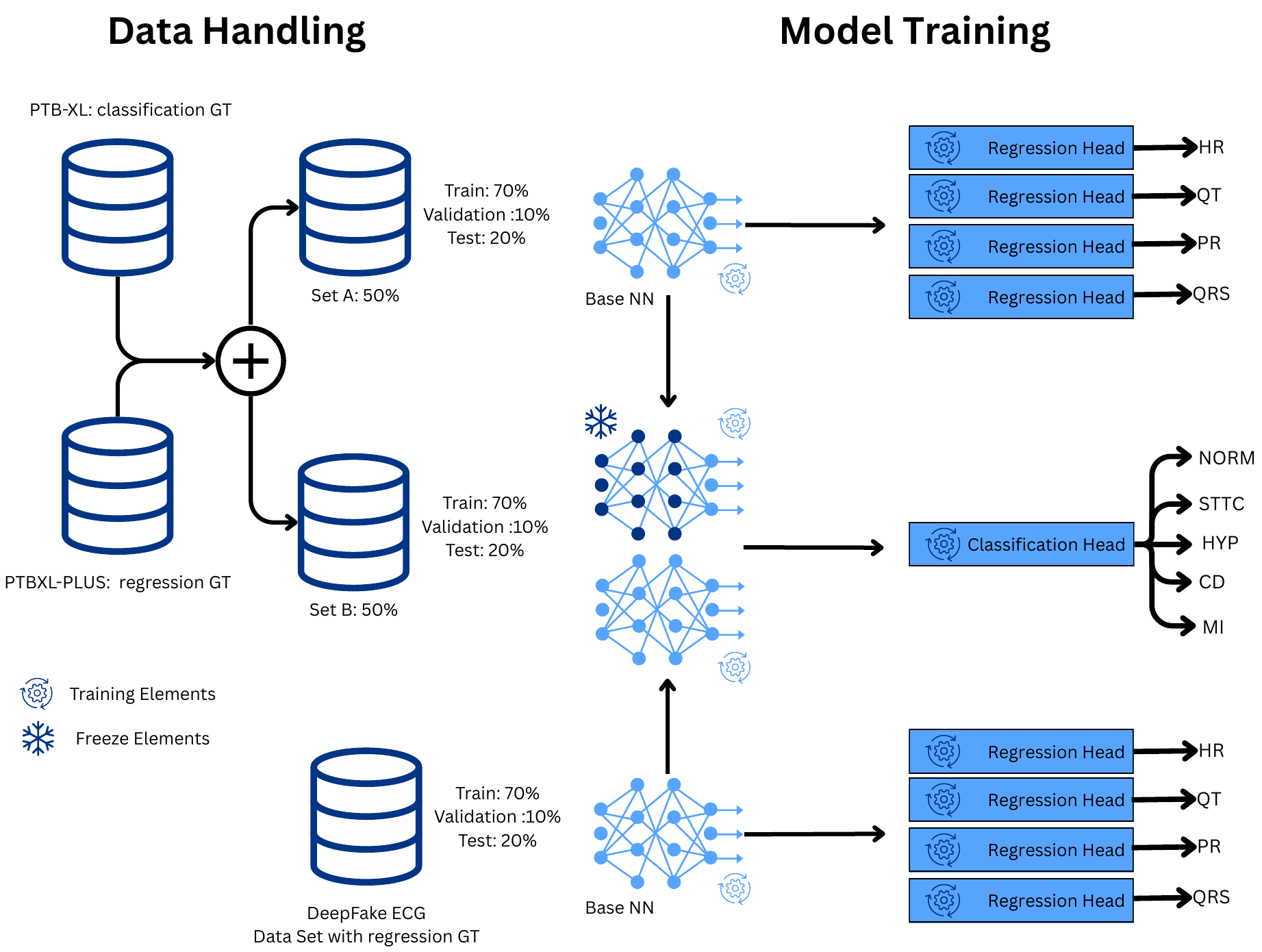}
    \caption{Overview of the methodology showing data handling from ECG datasets and the training pipeline involving regression and classification models with transfer learning.}
    \label{fig:figure2_overall-diagram}
\end{figure}

Figure~\ref{fig:figure2_overall-diagram} provides a clearer outline of the methodology described in this section. The diagram represents the data flow and experimental pipeline used in this study to test inductive transfer learning. 


The PTB-XL dataset~\cite{i3} includes real ECG signals annotated with classification labels for five diagnostic categories, such as NORM, CD, STTC, HYP, and MI. The PTB-XL Plus dataset extends this by incorporating regression labels for key ECG parameters, including Heart Rate (HR), PR interval, QT interval, and QRS complex—alongside the same ECG signals. To conduct the experiments, the PTB-XL dataset was split into two subsets, Set A ($50\%$ from the full dataset) and Set B (the other $50\%$), ensuring a balanced class distribution across the five diagnostic categories. Set A was primarily designated for regression tasks, while Set B was reserved for classification tasks. Each set was further divided into training ($70\%$), validation ($10\%$), and test ($20\%$) subsets. In addition to PTB-XL datasets, the DeepFake ECG dataset~\cite{b1} was used to train regression models to test inductive transfer learning from synthetic to real, from regression to classification. 

The pipeline progresses from the data, split into two key tasks: classification and regression. The classification task categorizes ECG signals into five diagnostic classes, and the regression task predicts the four ECG parameters.

Finally, the transfer learning phase adapts the knowledge gained from regression tasks to enhance classification performance. This phase evaluates the transfer learning process under two settings, as shown in Figure~\ref {fig:figure2_overall-diagram}: from PTB-XL Set A to PTB-XL Set B (real-to-real data transfer) and from the DeepFake ECG dataset to PTB-XL Set B (synthetic-to-real data transfer). Apart from using these two different settings transfer learning phase, it also has two different modes of learning, namely freezing the initial layers of the neural network and not freezing the initial layers. That is represented in figure~\ref{fig:figure2_overall-diagram} as training elements and freezing elements of the neural network.

\subsection{Data Handling}

In this study, we utilized three datasets: the PTB-XL dataset~\cite{i3}, the PTB-XL Plus dataset~\cite{ptbxlplus1, ptbxlplus2}, and the Deepfake ECG dataset~\cite{b1}. The PTB-XL Plus dataset was integrated with the PTB-XL dataset to get ECG features similar to the features in Deepfake ECGs. The details of these datasets are discussed in the following sections. 

\subsubsection{PTB-XL and PTB-XL Plus dataset}
The PTB-XL dataset~\cite{i3} includes $21,837$ clinical 12-lead ECG records from $18,885$ patients, each lasting for 10 seconds and sampled at $500 Hz$. The records are annotated with diagnostic statements classified into five main categories: NORM (Normal ECG), CD (Conduction Disturbance), STTC (ST/T change), MI (Myocardial Infarction), and HYP (Hypertrophy). 

From the 12 leads, the following eight leads are used for the experiments: Lead I (Index 0), Lead II (Index 1), V1 (Index 6), V2 (Index 7), V3 (Index 8), V4 (Index 9), V5 (Index 10), V6 (Index 11), in order to be compatible with the synthetically generated ECG dataset in the transfer learning process. 

\begin{figure}[h!]
    \centering
    \includegraphics[width=0.9\textwidth]{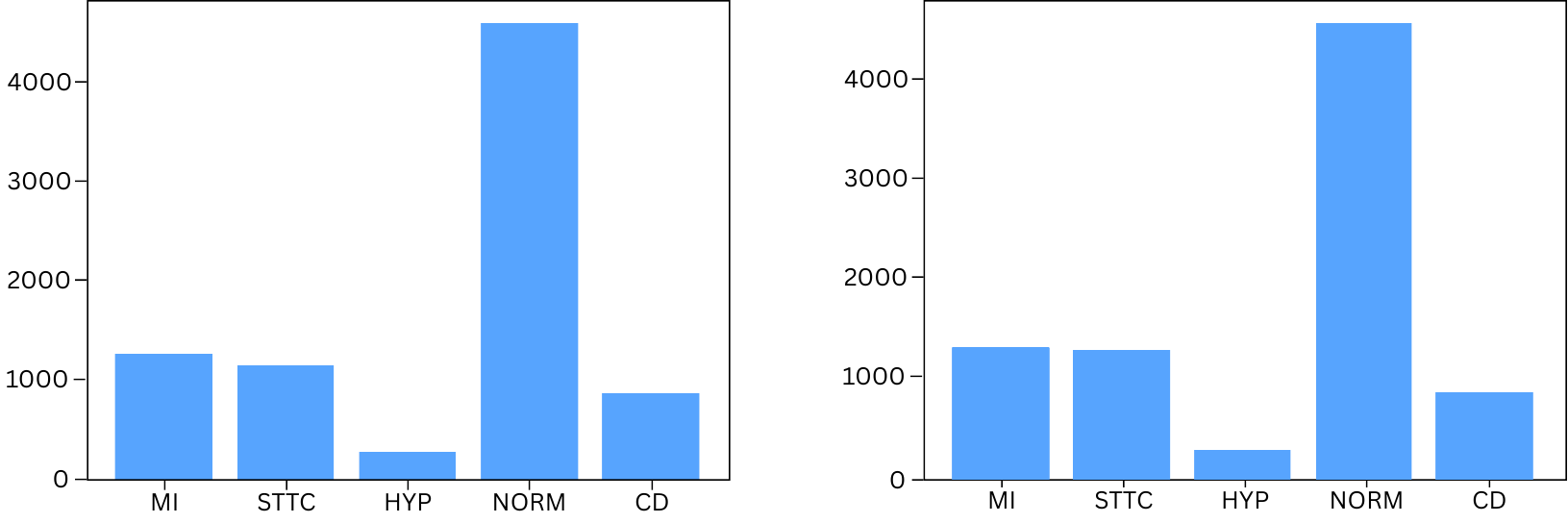}
    \caption{Distribution of ECG samples across the five diagnostic classes (NORM, CD, STTC, MI, and HYP) in PTB-XL Set A and Set B.}
    \label{fig:figure3_distribution}
\end{figure}

       
      


The PTB-XL dataset was preprocessed and cleaned using exploratory data analysis techniques. Further, we have filtered the multi-class labeled ECG signals, normalized each lead by scaling the values between 0 and 1, and only used ECG signals with a single class label from the 5-class classification set in order to make sure a given ECG signal falls only into a given class of the 5-class classification. The resulting dataset was divided mainly into two separate datasets, namely PTB-XL-A and PTB-XL-B, randomly but ensuring a fair class distribution of the five classes in sets A and B. This is in order to test the transfer learning techniques' effectiveness from real ECG to real ECG.  Figure~\ref{fig:figure3_distribution}  shows the class distribution of the five classes in sets A and B, where the x-axis denotes the five classes and the y-axis denotes the number of ECG samples. Each of these sets was split into train($70\%$), validation($10\%$), and test($20\%$). The dataset was output as an $8×5000$ matrix, and the classification labels were provided as a one-hot encoded list for five-class classification.

The PTB-XL Plus dataset~\cite{ptbxlplus1, ptbxlplus2}  is an enhanced version of the PTB-XL dataset, specifically curated to include regression values for key ECG parameters, including Heart Rate (HR), PR interval, QT interval, and QRS complex of our interest. This augmentation addresses the limitation of the original PTB-XL dataset, which only provided classification labels. For this research, the PTB-XL Plus dataset was used extensively for regression tasks and transfer learning experiments. It provided the required ground truth values for training regression models. As the only difference between the PTB-XL dataset and the PTB-XL Plus dataset is the regression labels, the data splits and output formats are exactly the same as the PTB-XL dataset that was mentioned in the above section.

\subsubsection{Deepfake ECG dataset}
\begin{figure}[ht]
\centering
\includegraphics[width=\linewidth]{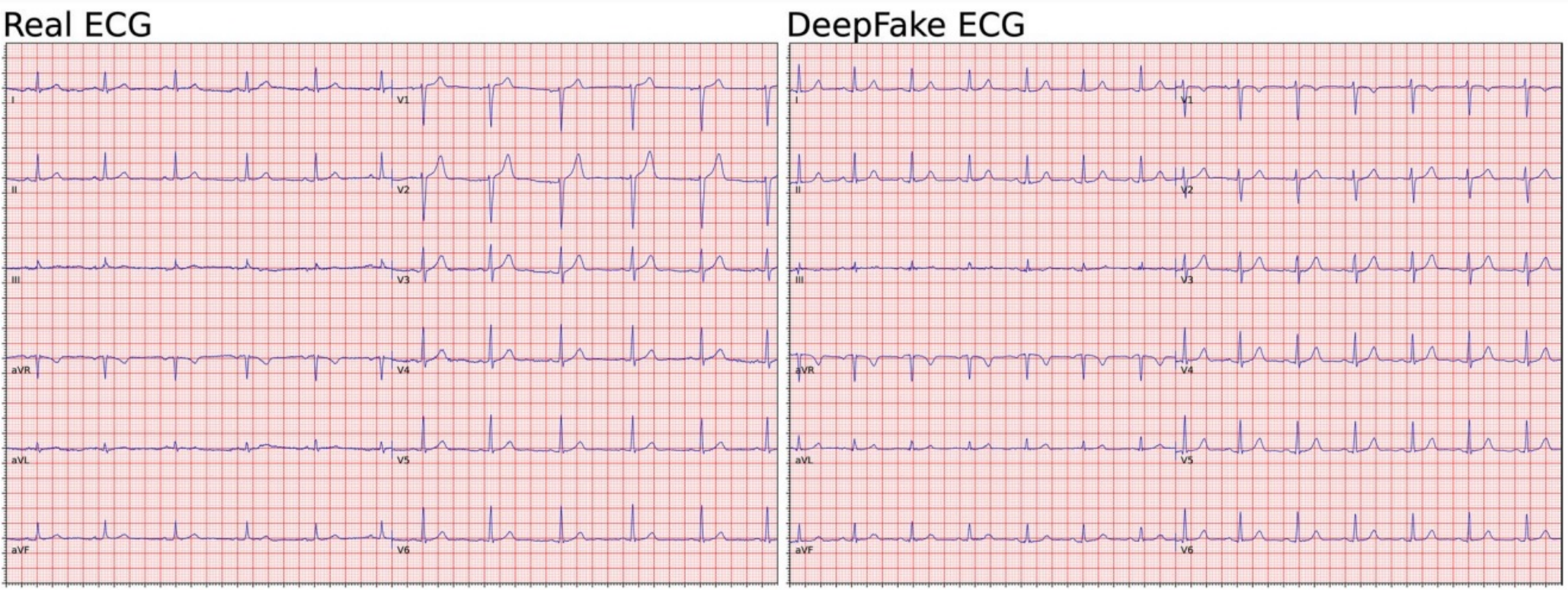}
\caption{Comparison of examples of a real ECG (left lane) and a DeepFake ECG (right lane).\cite{b1}}
\label{fig:deepfake_vs_real_ecg}
\end{figure}

\hfill\\  \indent \indent 
The DeepFake ECG dataset~\cite{b1} is a synthetically generated dataset designed to mimic real human ECG signals, addressing privacy concerns and the need for large training datasets in medical research. Developed using Generative Adversarial Networks (GANs)~\cite{r_gan}, this dataset provides ECG records that closely resemble the statistical properties and dynamics of real ECG signals. It has 121977 ECG signals along with corresponding labels, where we are considering only the four key parameters: Heart Rate (HR), PR interval, QT interval, and QRS complex. Each signal consists of 8-lead values [lead names from first column to eighth column: 'I', 'II', 'V1', 'V2', 'V3', 'V4', 'V5', 'V6'] for 10s (5000 values per lead). The signals for each lead were normalized by scaling their values between 0 and 1. The final output was an 8×5000 matrix, with regression labels provided for one parameter at a time. Figure \ref{fig:deepfake_vs_real_ecg} shows a comparison of the 8 leads of real ECG and corresponding 8 leads of DeepFake ECG. 

\subsection{Model Training and evaluation}

\subsubsection{Regression Task}
For the regression task, subset A of the PTB-XL Plus dataset and the deepfake dataset was used. Both of the datasets were separately used to train the models on regression and then used as the base models for the transfer learning process. The raw ECG data from both real and synthetic ECG datasets were preprocessed to ensure consistency and quality. This involved normalizing each lead by scaling the values between 0 and 1 and splitting the data into training, validation, and test sets. We utilized separate deep learning models to predict each of the four key ECG parameters: Heart Rate (HR), PR interval, QT interval, and QRS complex. The models were designed using the same 1D-CNN model used in classification tasks, but the final layer was adjusted for regression tasks.

\begin{figure}[h!]
    \centering
    \includegraphics[width=1\textwidth]{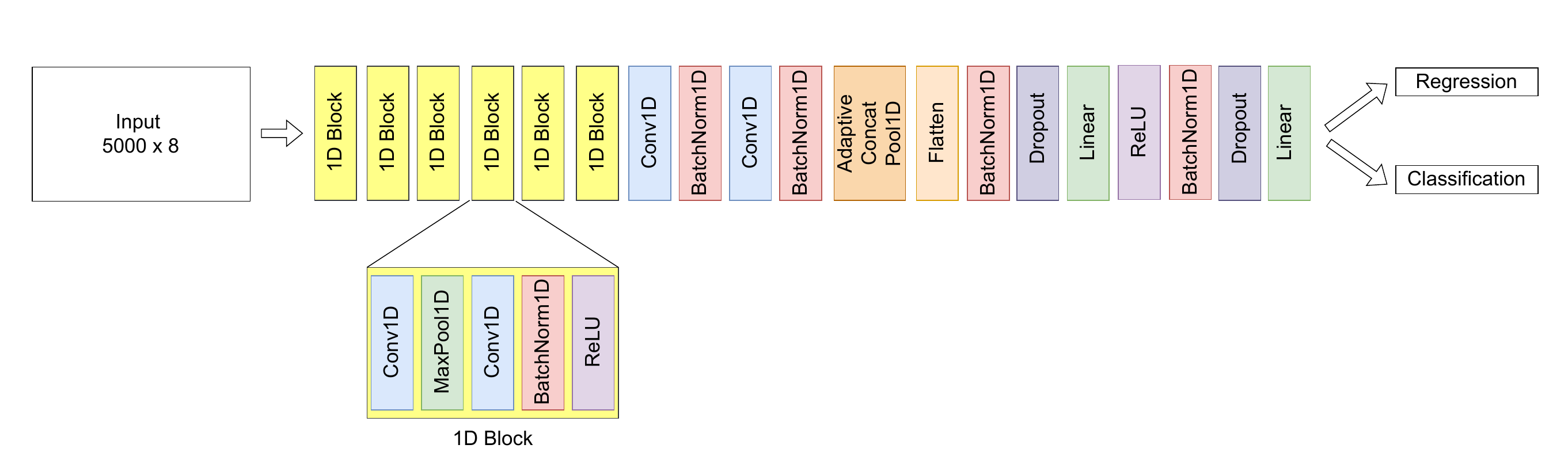} 
    \caption{ 1D-CNN Model for Regression and Classification. Input is from 8 leads and 5000 data points per lead.}
    \label{fig:1d_cnn_model_for_regression}
\end{figure}

The regression model, as shown in Figure \ref{fig:1d_cnn_model_for_regression}, employs a modified version of the 1D-CNN architecture used for classification, designed to predict four key ECG parameters: Heart Rate (HR), PR interval, QT interval, and QRS complex. Similar to the classification model, it utilizes multiple 1D blocks, each comprising Conv1D layers for extracting temporal features from the ECG signals, followed by max pooling layers to downsample the data. Batch normalization and ReLU activation layers ensure stable and efficient learning. However, for regression tasks, the model's final layer is adapted to output continuous values instead of class labels. This is done by replacing the classification layers with linear layers that generate scalar outputs for each of the four target ECG parameters. The architecture thus leverages CNN's feature extraction capabilities while catering specifically to the needs of regression tasks in ECG analysis.

Model training was performed using the Adam optimizer with an L1 loss function from the PyTorch nn library to minimize the error between predicted and true labels. ExponentialLR from the torch optim library was used as the learning rate scheduler to dynamically adjust the learning rate, with a patience of 50 epochs. The starting learning rate was set to 0.01, and the gamma value, which is set to control the exponential decay, was set to 0.99.
    
The performance of the regression models was evaluated using the MAE metric on both the training and validation datasets. Lower MAE values indicated better model performance. The results from these regression models were crucial for the subsequent transfer learning phase, where the learned features were adapted for classification tasks.

\subsubsection{ECG Classification}

For the classification task, subset B of the PTB-XL dataset was used, and after testing several deep learning models (model selection results summarized in the Results section), we employed a deep learning model with strengths of CNNs~\cite{i6, i7} but for 1-Dimensional input. We call the model 1D-CNN, where CNNs were utilized for their ability to capture local patterns and features in the ECG signals through convolutional layers. Specifically, 1D CNNs were used to process the temporal sequences of ECG data, as each single lead of the 8-lead ECG signal is for a duration of 10 seconds with 5000 values. With a batch size of 32, the input to the model has the shape of [32,8,5000]. The deeper layers of the complex model help to focus on different parts of the input sequence, effectively capturing complex dependencies and improving classification accuracy. The complex architecture leverages different CNN layers for feature extraction and sequence modeling, resulting in a robust model capable of classifying ECG signals into the five diagnostic categories. 

The 1D block, as shown in Figure \ref{fig:1d_cnn_model_for_regression}, is a key component of the 1D-CNN architecture used for ECG classification. It consists of a sequence of layers designed to extract and process features from the input ECG signals. The block begins with a 1D convolutional layer (Conv1D) that captures local patterns from the input data. This is followed by a 1D max pooling layer (MaxPool1D) to reduce the spatial dimensions and focus on the most prominent features. A second Conv1D layer is then applied for deeper feature extraction, followed by a batch normalization layer (BatchNorm1D) to stabilize the learning process. The block concludes with a ReLU activation function, adding non-linearity to the model and enhancing its ability to capture complex patterns.

The 1D-CNN model, illustrated in Figure \ref{fig:1d_cnn_model_for_regression}, consists of multiple 1D blocks, which are followed by additional Conv1D and batch normalization layers to further refine the extracted features. After the convolutional layers, the output is flattened and passed through fully connected layers (MLP) for final classification. ReLU activations and dropout layers are used to prevent overfitting and ensure generalization. The final output layer generates predictions for the five diagnostic categories: NORM, STTC, CD, HYP, and MI.

The final output layer used a softmax activation function to classify the ECG signals into one of the five diagnostic categories. The softmax function is defined as \eqref{eq:softmax}.

        \begin{equation}
    \text{softmax}(\mathbf{x})_i = \frac{\exp(x_i)}{\sum_{j} \exp(x_j)} \label{eq:softmax}
    \end{equation}

Model training was performed using the Adam optimizer with a categorical cross-entropy loss function to minimize the error between predicted and true labels. ExponentialLR from the torch optim library was used as the learning rate scheduler, with a patience of 50 epochs. Starting learning rate was set to 0.01 and the gamma value, which is set to control the exponential decay was set to 0.99.

Both Accuracy values and the Area Under the Curve (AUC)~\cite{rid_auc} values were measured and the performance of the model was evaluated using AUC  metric, which measures the model's ability to distinguish between classes, where higher AUC values indicating better performance.

\subsubsection{Inductive Transfer Learning}


Transfer learning was used. Two approaches were explored based on how the pre-trained model layers were treated during training; one involved freezing certain layers to retain the learned features, while the other allowed all layers to be retrained for the classification task. The transfer learning process involved adapting pre-trained regression models to the classification task by modifying the last layer of the model to output class probabilities instead of continuous values. The last layer of the model was replaced with a fully connected layer appropriate for the number of disease classes(5) in the classification task. Since we trained regression models for four different ECG parameters, we conducted separate experiments using each of them. Two approaches were explored based on how the pretrained model layers were treated during training, one involved freezing certain layers to retain the learned features, while the other allowed all layers to be retrained for the classification task. 

In the first setup, the first seven layers of the pre-trained regression model were frozen to retain the knowledge acquired during the regression task. This was achieved by setting the requires\_grad attribute to False for the parameters of the first seven layers. Freezing these layers helped to preserve the ECG-specific feature extraction capabilities while allowing the remaining layers to adapt to the classification task. In the second setup, none of the layers were frozen, allowing the model to update all weights during the classification task training. This approach aimed to fully optimize the model for disease classification by fine-tuning both the feature extraction and classification layers. 

To demonstrate the effectiveness of the transfer learning approach, the process was initially tested by transferring knowledge from the PTB-XL-A dataset to the PTB-XL-B dataset. After obtaining successful results from PTB-XL-A to PTB-XL-B transfer, the process was further tested by transferring knowledge from the DeepFake ECG dataset to the PTB-XL-B dataset to evaluate the effectiveness of using synthetic data for training and real data for testing.

With four different parameters and two fine-tuning strategies, we conducted a total of eight transfer learning experiments per dataset. Given that the model was trained on two datasets, this resulted in a total of sixteen experiments of transfer learning.

The model was trained using the Adam optimizer, with categorical cross-entropy as the loss function to reduce the error between predicted and actual labels. To adjust the learning rate over time, the ExponentialLR scheduler from the PyTorch \emph{optim} library was used. Training started with a learning rate of $0.01$, and the rate decayed exponentially with a gamma factor of $0.99$. Additionally, the scheduler was configured with a patience of $50$ epoch

The performance of the transfer learning models was evaluated using the Area Under the Curve (AUC) metric, comparing the results with standalone classification models to assess the effectiveness of transfer learning.

\section{Results}




 
The results section is divided into three parts, which are dedicated to model selection results, transfer learning between subsets of the PTB-XL dataset, and transfer learning from the DeepFake dataset to the PTB-XL dataset.

\subsection{Model Selection}
The first set of experiments was run to choose the best model for both the regression and classification of ECG signals. Various models were evaluated, namely RNN, LSTM, 2D-CNN, ViT and 1D-CNN. Table \ref{tab:regression_model_selection_results} presents the performance of different model architectures for regression on four parameters, namely HR parameter, QRS parameter, PR parameter, and QT parameter. Mean absolute error (MAE) for both the training and validation sets was used to compare the models' performance, as shown in Equation \eqref{eq:mae}. The best-performing model results for each parameter are highlighted in bold in Table  \ref{tab:regression_model_selection_results}. Here, \(n\) represents the total number of data points, \(y_i\) is the actual (true) value for the \(i\)-th data point, \(\hat{y}_i\) is the predicted value for the \(i\)-th data point, and \(\left| y_i - \hat{y}_i \right|\) is the absolute error for the \(i\)-th data point.

\begin{equation}
\text{MAE} = \frac{1}{n} \sum_{i=1}^{n} \left| y_i - \hat{y}_i \right| \label{eq:mae}
\end{equation}

Among the models, the 1D-CNN demonstrated superior performance across most parameters, with train MAE and val MAE values of 1.237 and 0.706 for HR, 3.259 and 3.007 for QRS, 5.801 and 5.110 for PR, and 6.351 and 4.130 for QT, respectively. Notably, ViT performed the worst in some areas, such as PR and QT predictions, indicating poor generalization to unseen data.

\renewcommand{\arraystretch}{1.2}

\setlength{\tabcolsep}{4pt}
\begin{table}[ht]
\centering
\begin{tabular}{|l|c|c|c|c|c|c|c|c|}
\hline
\multirow{2}{*}{Model} & \multicolumn{2}{c|}{HR} & \multicolumn{2}{c|}{QRS} & \multicolumn{2}{c|}{PR} & \multicolumn{2}{c|}{QT} \\
\cline{2-9}
 & Tr MAE(↓) & Val MAE(↓) & Tr MAE(↓) & Val MAE(↓) & Tr MAE(↓) & Val MAE(↓) & Tr MAE(↓) & Val MAE(↓)\\
\hline
RNN & 5.569 & 5.530 & 6.983 & 6.998 & 13.489 & 13.403 & 16.243 & 16.494 \\
\hline
LSTM & 5.536 & 5.546 & 6.819 & 6.896 & 13.431 & 13.338 & 13.536 & 13.892 \\
\hline
2D-CNN & 5.875 & 5.870 & 7.002 & 6.872 & 13.496 & 13.418 & 16.239 & 16.378 \\
\hline
ViT & 5.975 & 5.869 & 7.979 & 7.930 & 23.819 & 24.012 & 17.320 & 16.817 \\
\hline
\textbf{1D-CNN} & \textbf{1.237} & \textbf{0.706} & \textbf{3.259} & \textbf{3.007} & \textbf{5.801} & \textbf{5.110} & \textbf{6.351} & \textbf{4.130} \\
\hline
\end{tabular}
\caption{Comparison of various regression models on ECG signal parameters. Best-performing model results for each parameter are highlighted in bold.}
\label{tab:regression_model_selection_results}
\end{table}

The same set of model architectures was tested for the classification of five classes in the PTB-XL dataset. Accuracy values and AUC values for both the training dataset and validation dataset are shown in Table \ref{tab:classification_model_selection_results}. The best-performing model results for each metric are highlighted in bold. The accuracy is calculated using Equation \eqref{eq:accuracy}, where \( n \) is the total number of data points, and \( n_{\text{correct}} \) is the number of correctly classified data points.

\begin{equation}
\text{Accuracy} = \frac{n_{\text{correct}}}{n} \label{eq:accuracy}
\end{equation}

The multi-class AUC is computed using the One-vs-Rest (OvR) approach, as described in Equation \eqref{eq:auc_combined}. In this approach, each class \( c \) is treated as the positive class, while all other classes are considered negative. For each class, the AUC is calculated by integrating the true positive rate (TPR) over the false positive rate (FPR), where \( t \) represents the decision threshold that varies between 0 and 1. This threshold \( t \) is used to compute different points on the ROC curve by adjusting the classification boundary. The overall multi-class AUC score is the average of these class-specific AUC values.

\begin{equation}
\text{AUC}_{\text{multi-class}} = \frac{1}{C} \sum_{c=1}^{C} \left( \int_{0}^{1} \text{TPR}_c(t) \, d(\text{FPR}_c(t)) \right) \label{eq:auc_combined}
\end{equation}

The final multi-class AUC score is the average of the AUC scores across all classes. Let \(C\) be the total number of classes.

The 1D-CNN model outperformed all other models, achieving a training accuracy of 0.795 and an AUC of 0.905. The validation accuracy and AUC were also high, at 0.776 and 0.902, respectively. Other models, such as RNN, LSTM, 2D-CNN, Transformer Encoder, and ViT, showed lower accuracy and AUC values, with the highest validation AUC among them being 0.770 for the LSTM model.

\renewcommand{\arraystretch}{1.2}

\begin{table}[ht]
\centering
\begin{tabular}{|l|c|c|c|c|}
\hline
\multirow{2}{*}{Model} & \multicolumn{2}{c|}{Train} & \multicolumn{2}{c|}{Validation} \\
\cline{2-5}
 & Accuracy(↑) & AUC(↑) & Accuracy(↑) & AUC(↑) \\
\hline
RNN & 0.559 & 0.758 & 0.557 & 0.736 \\
\hline
LSTM & 0.560 & 0.779 & 0.551 & 0.770 \\
\hline
2D-CNN & 0.560 & 0.775 & 0.551 & 0.769 \\
\hline
ViT & 0.560 & 0.776 & 0.551 & 0.769 \\
\hline
\textbf{1D-CNN} & \textbf{0.795} & \textbf{0.905} & \textbf{0.776} & \textbf{0.902} \\
\hline
\end{tabular}
\caption{Comparison of various classification models on five-class ECG classification on PTB-XL dataset. Best performing model results for each metric are highlighted in bold.}
\label{tab:classification_model_selection_results}
\end{table}

\subsection{Inductive Transfer Learning from Real to Real data}
Following the selection of the 1D-CNN model as the optimal architecture for both regression and classification tasks, all subsequent experiments were conducted using this model. Initially, it was necessary to establish benchmark results for the classification task without using transfer learning techniques, and those results are presented in the first row of Table \ref{tab:transfer_learning_with_in_PTB_XL}. The remaining results in Table \ref{tab:transfer_learning_with_in_PTB_XL} are divided into two parts: transfer learning using models with the first block frozen and transfer learning using models without frozen layers. The best-performing results for accuracy and AUC in each setting are highlighted in bold in Table \ref{tab:transfer_learning_with_in_PTB_XL}. In the table, the parameter (HR, QRS, PR, QT) indicates the specific ECG parameter predicted in the regression task, which was used for initial training before applying transfer learning.(see Methods) 

For the frozen condition, the transfer learning models showed improvements over the baseline in both training and validation, as demonstrated by the increased accuracy and AUC scores across most metrics. Notably, the best transfer performance was achieved using the QT interval model with an AUC of 0.929 in the training set, 0.890 in the validation set, and 0.891 in the test set, compared to the baseline AUCs of 0.905, 0.903, and 0.884, respectively. The transfer learning models using HR and QRS labels also demonstrated competitive results, particularly on the validation and test sets, where QRS achieved the highest test set accuracy of 0.782 with an AUC of 0.898.

When no layers were frozen, further improvements were observed. The best performing model was again based on the QT interval, achieving the highest AUC of 0.949 on the training set, 0.901 on the validation set, and 0.893 on the test set. The transfer learning models for the HR and QRS labels similarly showed strong performance, with HR achieving an AUC of 0.892 on the test set and QRS yielding the highest test AUC of 0.906.

\renewcommand{\arraystretch}{1.2}

\begin{table}[ht]
\centering
\begin{tabular}{|c|c|c|c|c|c|c|c|}
\hline
\multirow{2}{*}{} & \multirow{2}{*}{} & \multicolumn{2}{c|}{Train} & \multicolumn{2}{c|}{Validation} & \multicolumn{2}{c|}{Test} \\
\cline{3-8}
 & & Accuracy(↑) & AUC(↑) & Accuracy(↑) & AUC(↑) & Accuracy(↑) & AUC(↑) \\
\hline
\multirow{1}{*}{} & Baseline & 0.795 & 0.905 & 0.776 & 0.903 & 0.775 & 0.884 \\
\hline
\multirow{4}{*}{\rotatebox[origin=c]{90}{1-7 Frozen}} & Transfer HR & 0.803 & 0.927 & 0.796 & \textbf{0.907} & 0.769 & 0.881 \\
\cline{2-8}
 & Transfer QRS & 0.815 & 0.925 & 0.781 & 0.883 & 0.782 & 0.898 \\
\cline{2-8}
 & Transfer PR & 0.785 & 0.910 & 0.765 & 0.902 & 0.775 & 0.875 \\
\cline{2-8}
 & Transfer QT & 0.806 & 0.929 & 0.754 & 0.890 & 0.759 & 0.891 \\
\hline
\multirow{4}{*}{\rotatebox[origin=c]{90}{Not Frozen}} & Transfer HR & 0.831 & 0.946 & 0.752 & 0.894 & \textbf{0.787} & 0.892 \\
\cline{2-8}
 & Transfer QRS & 0.802 & 0.915 & 0.791 & 0.892 & 0.785 & \textbf{0.906} \\
\cline{2-8}
 & Transfer PR & 0.812 & 0.924 & 0.785 & 0.902 & 0.766 & 0.883 \\
\cline{2-8}
 & Transfer QT & \textbf{0.838} & \textbf{0.949} & \textbf{0.808} & 0.901 & 0.783 & 0.893 \\
\hline
\end{tabular}
\caption{Inductive transfer learning results comparison using real data of PTB-XL dataset. Baseline results are from a model trained for a classification task using the training data of the classification subset of PTB-XL. Best performing results for accuracy and AUC in each setting are highlighted in bold.}
\label{tab:transfer_learning_with_in_PTB_XL}
\end{table} 

\subsection{Inductive Transfer Learning from Synthetic to Real data}
Similar to the previous methodology, the 1D-CNN model was pretrained on synthetic regression data from the deepfake dataset and then transferred this pre-trained model to classify real-world data from the PTB-XL dataset. Similar to our previous approach, we used two strategies: freezing the first seven layers of the network and not freezing any layers during the transfer learning process. The results, measured in terms of accuracy and area under the curve (AUC), are presented in the table \ref{tab:transfer_learning_from_deepfake} and the best-performing results for accuracy and AUC in each setting are highlighted in bold.

For the frozen condition, the QT transfer model yielded a solid training performance with 0.830 accuracy and 0.940 AUC. However, the validation and test performances were poor, with accuracies of 0.761 and 0.771, and AUC values of 0.897 and 0.882, respectively. Both QRS and HR transfer models have done well in the validation dataset with 0.908 and 0.904 AUC values, but both models do not show similar improvements in the Test dataset. All the frozen models have done poorly in the test dataset with lower accuracy and AUC values compared to the baseline results.

Regarding the transfer learning with no frozen layers, similar to the frozen condition, QT transfer model led to strong training results (0.806 accuracy and 0.912 AUC), with validation accuracy of 0.765 and AUC of 0.913. Test performance was also robust, with an accuracy of 0.778 and AUC of 0.895. PR transfer model’s accuracy on the training set was relatively lower (0.761), and validation accuracy also remained low at 0.754. However, the test accuracy reached 0.764, with the test AUC notably increasing to 0.888. Overall, all models except for the QRS model have shown higher test AUC values compared to the baseline results.

\renewcommand{\arraystretch}{1.2}

\begin{table}[ht]
\centering
\begin{tabular}{|c|c|c|c|c|c|c|c|}
\hline
\multirow{2}{*}{} & \multirow{2}{*}{} & \multicolumn{2}{c|}{Train} & \multicolumn{2}{c|}{Validation} & \multicolumn{2}{c|}{Test} \\
\cline{3-8}
 & & Accuracy(↑) & AUC(↑) & Accuracy(↑) & AUC(↑) & Accuracy(↑) & AUC(↑) \\
\hline
\multirow{1}{*}{} & Baseline & 0.795 & 0.905 & 0.776 & 0.903 & 0.775 & 0.884 \\
\hline
\multirow{4}{*}{\rotatebox[origin=c]{90}{1-7 Frozen}} & Transfer HR & \textbf{0.835} & \textbf{0.949} & 0.781 & 0.904 & 0.771 & 0.880 \\
\cline{2-8}
 & Transfer QRS & 0.804 & 0.918 & \textbf{0.791} & 0.908 & 0.761 & 0.862 \\
\cline{2-8}
 & Transfer PR & 0.825 & 0.935 & 0.762 & 0.875 & 0.762 & 0.882 \\
\cline{2-8}
 & Transfer QT & 0.830 & 0.940 & 0.761 & 0.897 & 0.771 & 0.882 \\
\hline
\multirow{4}{*}{\rotatebox[origin=c]{90}{Not Frozen}} & Transfer HR & 0.801 & 0.904 & 0.785 & 0.906 & 0.764 & 0.887 \\
\cline{2-8}
 & Transfer QRS & 0.822 & 0.933 & 0.777 & 0.897 & \textbf{0.783} & 0.876 \\
\cline{2-8}
 & Transfer PR & 0.761 & 0.865 & 0.754 & 0.886 & 0.764 & 0.888 \\
\cline{2-8}
 & Transfer QT & 0.806 & 0.912 & 0.765 & \textbf{0.913} & 0.778 & \textbf{0.895} \\
\hline
\end{tabular}
\caption{Baseline Results and Transfer Learning Results between Deepfake dataset and PTB-XL dataset. Best performing results for accuracy and AUC in each setting are highlighted in bold}
\label{tab:transfer_learning_from_deepfake}
\end{table}

\section{Discussion}
In this study, we evaluated several deep learning models for regression and classification tasks on ECG data from the DeepFake dataset and PTB-XL dataset, followed by transfer learning experiments within the PTB-XL dataset and from the DeepFake dataset to the PTB-XL dataset. Our results highlight the effectiveness of 1D-CNN models, as well as the potential benefits and challenges associated with transfer learning across ECG datasets.

The performance of various models for both regression and classification tasks revealed that the 1D-CNN outperformed alternative architectures like RNN, LSTM, 2D-CNN, and ViT. For the regression task, the 1D-CNN achieved the lowest mean absolute error (MAE) across all ECG parameters, particularly excelling in the HR and QT intervals. In contrast, models like RNN and LSTM showed higher MAE values, demonstrating their lower suitability for these tasks. The superior performance of the 1D-CNN aligns with its design, which leverages convolutional operations to capture temporal dependencies within the one-dimensional ECG signals effectively, leading to better generalization and lower errors in both training and validation datasets. Similarly, for classification tasks, the 1D-CNN demonstrated the highest accuracy and AUC, surpassing all other models. Both of these observations suggest that 1D-CNN is the most suitable model architecture among the tested model architectures for analysing ECG signals.

Following model selection, we applied transfer learning techniques within subsets of the PTB-XL dataset using the 1D-CNN model. In the approach where the first seven layers were frozen, models which was pretrained on the QT interval gave higher accuracy values and AUC values across all train, validation, and test datasets. The model, which was pretrained on QRS interval, also gave higher test AUC results compared to baseline results. 

Further improvements were observed when all layers were trainable, with the QRS model achieving the highest test AUC of 0.906 of all the results. The QT transfer model gave highly consistent accuracy and AUC results across the train, validation, and test datasets.

It can be noted that QT and QRS models have given good accuracy and AUC values compared to the baseline across both frozen and unfrozen approaches. This demonstrates that transfer learning from one ECG parameter to another within the same dataset can enhance model performance, likely due to the shared underlying physiological information between different ECG parameters.

As the next phase of the research, transfer learning experiments from the DeepFake dataset to PTB-XL were run, and they provided mixed results. Regarding the frozen condition, transfer learning improved the performance of models on the training set, but test performance was generally lower when compared to the baseline.

When no layers were frozen during transfer learning from synthetic to real data, several models demonstrated promising results. The QT transfer model achieved particularly strong performance with a test AUC of 0.895, surpassing the baseline AUC of 0.884. This improvement suggests that knowledge transfer from synthetic to real ECG data is not only feasible but can enhance model performance when properly implemented. The PR and HR transfer models also showed competitive results, with test AUCs of 0.888 and 0.887, respectively, further supporting the viability of using synthetic data for pre-training. While there is still room for improvement, these results demonstrate that synthetic ECG data can serve as a valuable resource for developing robust classification models, particularly in scenarios where access to real patient data is limited due to privacy concerns. The success of transfer learning from synthetic to real data points to a promising direction for future research in ECG analysis. These findings suggest that continued refinement of synthetic data generation techniques, coupled with advanced transfer learning strategies, could lead to even more significant improvements in model performance while addressing critical privacy concerns in medical data analysis. 

\section{Conclusion and future work}

The exploration of transfer learning within ECG datasets revealed clear advantages. Inductive transfer learning across subsets of the PTB-XL dataset, particularly when using the QT and QRS parameters, demonstrated notable improvements over baseline models, with significant gains in accuracy and AUC values. Both frozen and non-frozen layer strategies highlighted the potential of shared physiological information to enhance model performance. The highest test AUC achieved by the QRS model shows the effectiveness of transfer learning for ECG analysis. Transfer learning experiments from the synthetic DeepFake dataset to PTB-XL yielded mixed results, and the success of certain models, such as the QT transfer model, highlights the potential of synthetic data for future applications.

The research successfully fulfilled the objectives of demonstrating the potential of inductive transfer learning techniques from regression to ECG classification tasks. These findings offer valuable insights for future studies and pave the way for more advanced applications of transfer learning in the field of cardiovascular signal processing, especially through the integration of synthetic data.

\section*{Data availability}
The datasets analyzed during the current study are available in the PhysioBank repository and Kaggle:
\begin{itemize}[nosep]
    \item PTB XL Dataset: \href{https://physionet.org/content/ptb-xl/1.0.3/}{https://physionet.org/content/ptb-xl/1.0.3/}
    \item PTB XL PLUS Dataset: \href{https://physionet.org/content/ptb-xl-plus/1.0.1/}{https://physionet.org/content/ptb-xl-plus/1.0.1/}
    \item Deepfake ECG Dataset: \href{https://www.kaggle.com/datasets/vlbthambawita/deepfake-ecg}{https://www.kaggle.com/datasets/vlbthambawita/deepfake-ecg}
\end{itemize}

\section*{Code Availability}
All the source code related to the experiments conducted in this study is available at our open-source GitHub to reproduce the results: \href{https://github.com/cepdnaclk/e18-4yp-GPU-Acceleration-for-Deep-Learning-based-Comprehensive-ECG-analysis}{repository}.

\section*{Author contributions}

V.T. and I.N. conceptualized the study, methodology and provided resources. I.F., A.F. and R.J. conducted the investigation, software, validation and wrote the original draft.  R.R., V.T., and I.N. provided supervision. All authors reviewed and approved the final manuscript.


\section*{Corresponding authors}
Correspondence to Ridma Jayasundara (ridmajayasundara@eng.pdn.ac.lk), Isuru Nawinne (isurunawinne@eng.pdn.ac.lk)

\section*{Competing interests}
The authors declare no competing interests.

\section*{Funding}
This study was not funded.

\bibliography{references}

\end{document}